\documentclass[twocolumn,runningheads]{svjour2}

\smartqed  

\usepackage{graphicx}

\usepackage[T1]{fontenc}
\usepackage[latin1]{inputenc}
\usepackage{color}
\usepackage{multirow}
\usepackage{supertabular}
\usepackage{amsmath} 

\begin{document} 

\journalname{International Journal of Advanced Manufacturing Technology}

\title{Self-excited vibrations in turning: cutting moment analysis}


\author{Olivier~Cahuc \and  Jean-Yves~K'nevez \and Alain~G\'erard \and Philippe~Darnis \and Ga\"etan~Albert \and  Claudiu~F.~Bisu \and C\'eline~G\'erard}

\institute{G. Albert  \and O. Cahuc \and  A. G\'erard (corresponding author) \and  J-Y. K'nevez \at
              Universit\'e de Bordeaux - CNRS UMR 5469,\\351 cours de la Lib\'eration, 33405 Talence-cedex France (EU) 
              \\\email{alain.gerard@u-bordeaux1.fr} 
\and P. Darnis \at Universit\'e de Bordeaux - IUT EA 496,\\15 rue Naudet, 33175 Gradignan Cedex France (EU) 
 \and C.F. Bisu \at University Politehnica from Bucharest,\\313 Splaiul Independentei,  060042 Bucharest Roumanie (EU)
              \and C. G\'erard \at Centre des Mat\'eriaux (MAT), CNRS : UMR 7633 - Mines ParisTech - BP 87 - 91003 EVRY CEDEX, France}

\date{Received:  / Accepted: }

\maketitle

\begin{abstract}

This work aims at analysing the moment effects at the tool tip point and at the central axis, in the framework of a turning process. A testing device in turning, including a six-component dynamometer, is used to measure the complete torsor of the cutting actions in the case of self-excited vibrations. Many results are obtained regarding the mechanical actions torsor. A confrontation of the moment components at the tool tip and at the central axis is carried out. It clearly appears that analysing moments at the central axis avoids the disturbances induced by the transport of the moment of the mechanical actions resultant at the tool tip point. For instance, the order relation between the components of the forces is single. Furthermore, the order relation between the moments components expressed at the tool tip point is also unic and the same one. But at the central axis, two different order relations regarding moments are conceivable. A modification in the rolling moment localization in the (y, z) tool plan is associated to these two order relations. Thus, the moments components at the central axis are particularly sensitive at the disturbances of machining, here the self-excited vibrations.


\end{abstract}

\section*{Nomenclature}
\label{sec:Nomenclature}

	\begin{tabular}{lp{6.cm}}
\raggedright 

A & Central axis point;\\
$[A]_{o}$ & Actions torsor exerted at the tool tip in O point;  \\
$a_{p}$  & Depth of cut;\\
\textbf{BT} & Block Tool\\
$f$  & Feed rate (mm/rev)\\
$F_{v}$  & Variable  cutting force (N);\\
$F_{n}$ & Nominal cutting force (N);\\
$F_{x}$ & Effort along cross direction (N); \\
$F_{y}$ & Effort along cutting axis (N); \\
$F_{z}$ & Effort along feed rate axis (N); \\
$\textbf{M}_{_{A}}$ & Cutting forces minimum moment in A that act on the tool (dN.m);\\
$\textbf{M}_{_{O}}$ & Cutting forces moment in the O point that act on the tool (dN.m); \\
$M_{ox}$ & Moment component that act at the tool tip along cross direction (dN.m); \\
$M_{oy}$ & Moment component that act at the tool tip along cutting axis (dN.m); \\
$M_{oz}$ & Moment component that act at the tool tip along feed rate axis (dN.m); \\
$M_{ax}$ & Moment component at the central axis along cross direction (dN.m); \\
$M_{ay}$ & Moment component at the central axis along cutting axis (dN.m); \\
$M_{az}$ & Moment component at the central axis along feed rate axis (dN.m); \\
$N$ & Rotational velocity (rpm); \\
O & Tool point reference; \\
O' & Dynamometer center transducer; \\
$\textbf{R}$ & Cutting forces vector sum which act at the tool (N); \\
$r_{\epsilon}$ & Cutting edge radius (mm); \\
R & Sharpness radius (mm); \\
\textbf{WTM} & Workpiece-Tool-Machine tool; \\
x & Radial direction; \\
y	&	Cutting axis; \\
z	&	Feed rate direction; \\

\end{tabular}

	\begin{tabular}{lp{6.8cm}}
\raggedright

$\alpha$	&	Clearance angle (degree); \\
$\gamma$	&	Cutting angle (degree); \\
$\lambda_{s}$	&	Inclination angle of tool edge (degree); \\
$\kappa_{r}$	&	Direction angle of the cutting edge (degree); \\

\end{tabular}

\section{Introduction}
\label{sec:1}

In the three-dimensional cutting configuration, the mechanical actions torsor (forces and moments), is often truncated: the moments part of this torsor is neglected due to the lack of adapted metrology \cite{mehdi-A-play-02b}, \cite{yaldiz-unsacar-06b},  \cite{yaldiz-unsacar-06a}, \cite{lian-A-huang-07}, \cite{marui-A-kato-83b}. 
Unfortunately, until now the results on the cutting forces are almost still validated using platforms of forces (dynamometers) measuring the forces three components. However, forces and pure moments (or torque) can be measured \cite{couetard-93}. Recently, an application consisting in six-component measurements of the actions torsor in 
cutting process was carried out in the case of high speed milling \cite{couetard-A-darnis-01}, drilling \cite{laporte-AA-darnis-09}, \cite{yaldiz-AA-isik-07}, etc. Cahuc et al., in \cite{cahuc-AA-battaglia-01}, present another use of this six-component dynamometer in an experimental study: the taking into account of the cutting moments allows a better machine tool power consumption evaluation. It allows a better approach of the cut \cite{bisu-AAAAA-ispas-07a},  \cite{couetard-00a}, \cite{laporte-AA-darnis-07} and thus allows to reach new properties of the vibrations of the chip-tool-workpiece system in the dynamic case.

Moreover, the tool torsor has the advantage of being transportable in any space point and, especially, at the tool tip, here called O point. The following study is carried out in several stages, including two major ones. The first one is related to the analysis of forces. The second one is dedicated to determining of the central axis and a first moments analysis at the central axis during the cut. 

In section~\ref{sec:2}, we present first the experimental device used and associated measurement elements. 

Section~\ref{sec:3} is devoted to the measurement of the cutting process torsor. An analysis of the forces exerted during the cuttig process is performed. It allows to establish experimentaly several properties of the cutting actions resultant. The case of the moments at the tool tip point is also accurately examinated. The central axis of the torsor is required (section~\ref{sec:CentralAxis}). The existence of central axes highlighted from the multiple tests confirms noticeably the presence of moments at the tool tip point. In section~\ref{sec:5} we carry out more particularly the analysis of the moments at the central axis by studying the case the most sensitive to vibrations (ap=5 mm, f = 0.1 mm/rev). Before concluding, this study gives a certain number of properties and drives to some innovative reflexions.

\section{Experimental device}
\label{sec:2}

\begin{figure}
	\centering
		\includegraphics[width=0.48\textwidth]{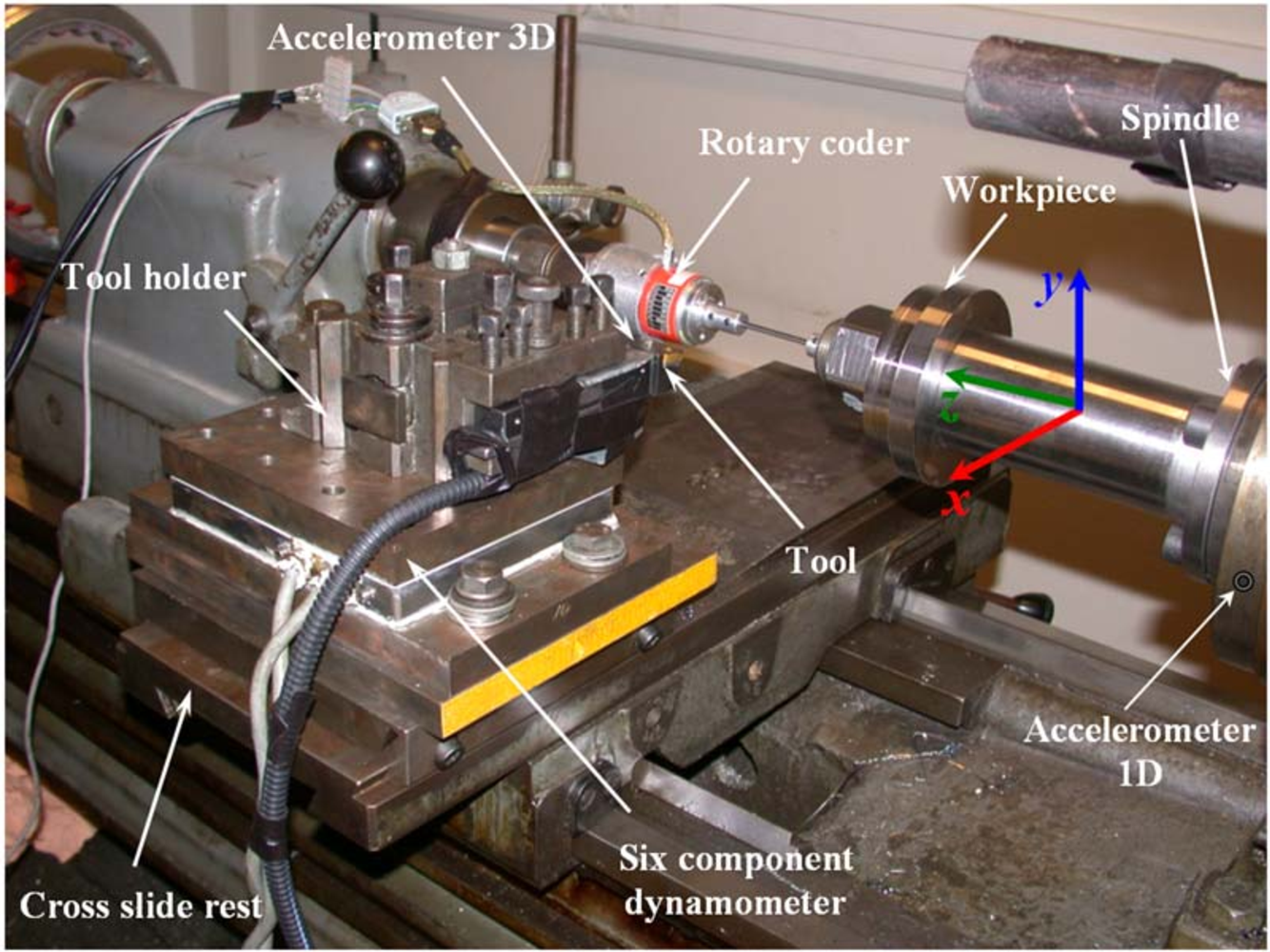}
	\caption{Experimental device and associated measurement elements}
	\label{fig1}
\end{figure}

\begin{figure}[htbp]
	\centering
		\includegraphics[width=0.48\textwidth]{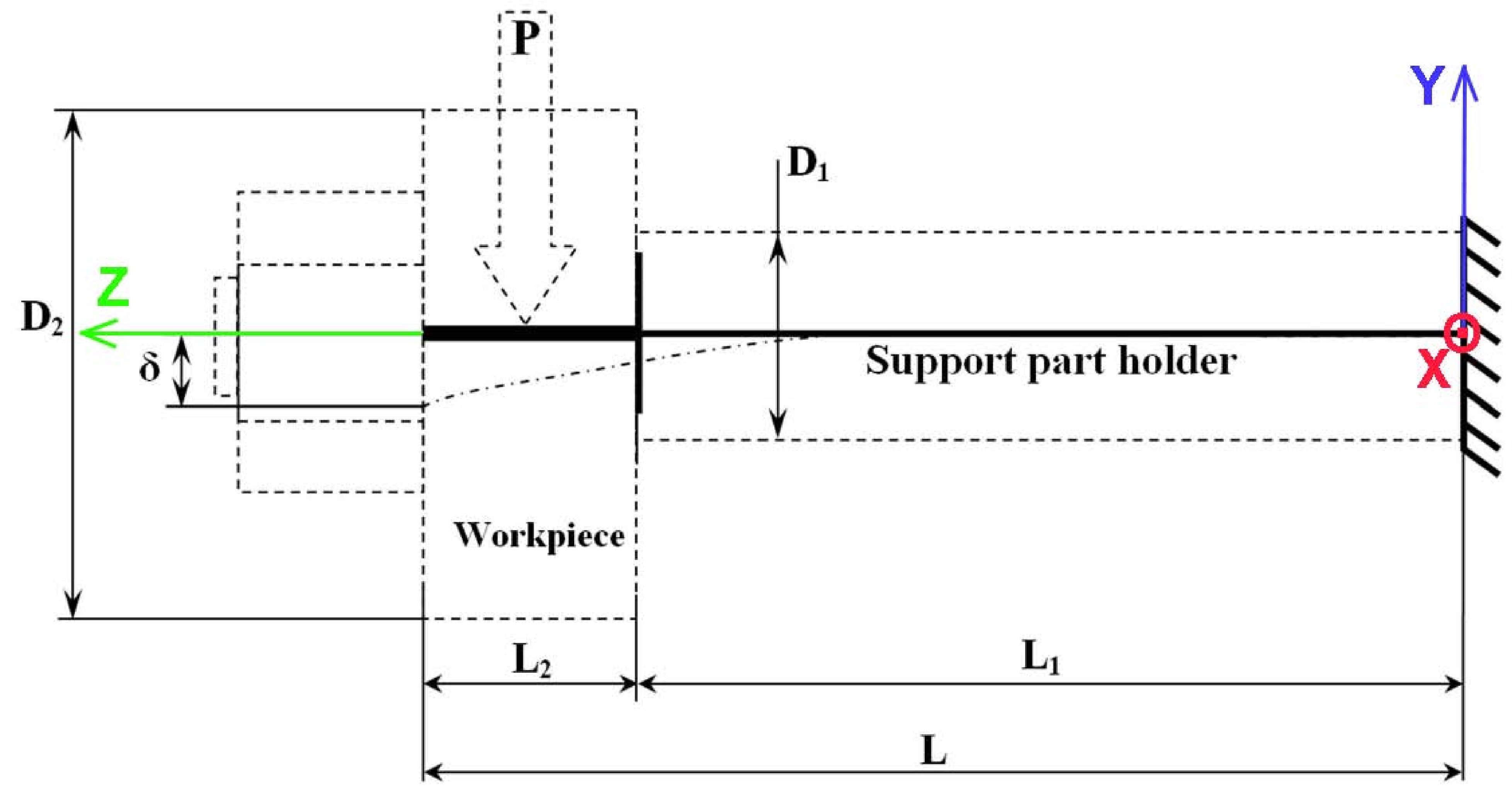}
	\caption{Geometry and dimension of the support part holder and the workpiece}
	\label{fig2}
\end{figure}

The experimental device presented in Fig.~\ref{fig1} is a conventional lathe (Ernault HN 400). The machining system dynamic behaviour is identified using a three-direction accelerometer fixed on the tool. Two unidirectional accelerometers are positioned on the lathe and on the front bearing of the spindle, in order to identify the influence of this one during the cutting process. All the cutting actions (forces and torques at the tool tip point) are measured by a six-component dynamometer \cite{couetard-93}. The instantaneous spindle speed is continuously controlled (with an accuracy of 1\%) by a rotary encoder directly coupled with the workpiece. The connection is carried out by a rigid steel wire. The test workpieces are cylindrical and have a diameter of 120 mm and a length of 30 mm. The geometry of the support part holder was optimized by a finite element analysis (using SAMCEF$^{\textregistered}$ software) in order to confer a maximum rigidity to the unit. This procedure is described in \cite{bisu-07}. Thus, under a load P = 1,000 N, for material having a Young modulus E = 21$\times$10$^{5}$ N/mm$^{2}$, the workpiece dimensions selected are: D$_{1}$ = 60 mm (diameter), L$_{1}$ = 180 mm (length), for a bending stiffness of 7$\times$10$^{7}$ N/m (Fig.~\ref{fig2}) \cite{bisu-AAA-gerard-09}. These values are within the higher limit of the stiffness zone acceptable for a conventional lathe \cite{ispas-AA-anghel-99}, \cite{koenigsberger-tlusty-70}, \cite{konig-A-lauer-schmaltz-97}.

\begin{figure}[htbp]
	\centering
		\includegraphics[width=0.40\textwidth]{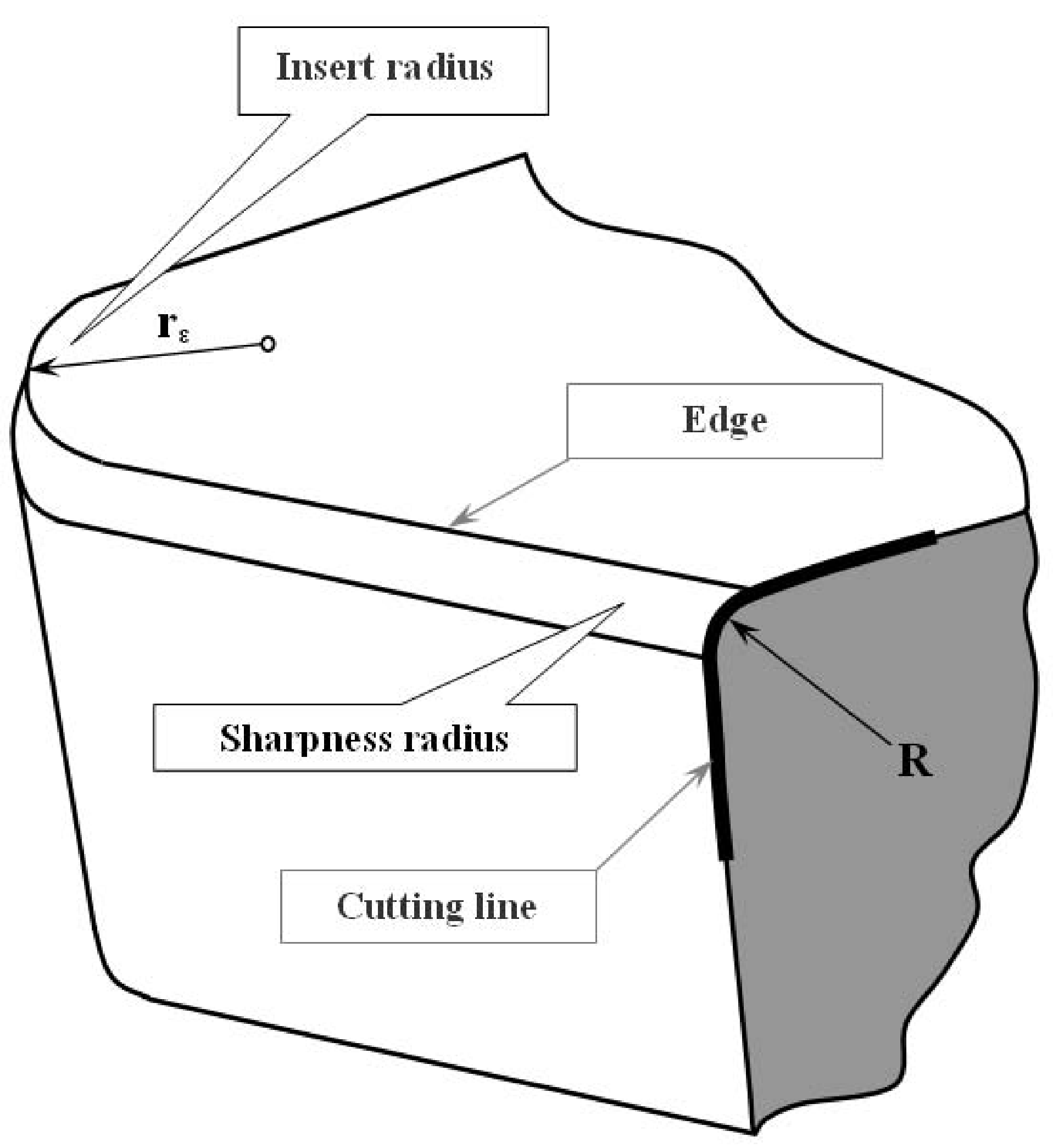}
	\caption{Tool geometry}
	\label{fig3}
\end{figure}

The insert tool used is a TNMA 16 04 12 type carbide not coated, without chip breaker. The machined material is a chrome molybdenum 42CrMo24 type alloy. Moreover, the tool geometry (Fig.~\ref{fig3}) is characterized by the cutting angle $\gamma$, the clearance angle $\alpha$, the inclination angle of edge $\gamma_{s}$, the direct angle $\kappa_{r}$, the nozzle radius $r_{\epsilon}$ and the sharpness radius $R$ \cite{laheurte-04}. In order to limit the influence of the tool wear on measurements, the tool insert is examined after each test, and is changed if necessary (Vb $\leq$ 0.2 mm ISO 3685). Nevertheless, the present study does not strictly speaking deal with fretting and tool wear. The tool parameters are detailed in the Table \ref{tabl-1}.

\begin{table}[htbp]
	\centering
		\begin{tabular}{|c|c|c|c|c|c|}
\hline
 $\gamma$ & $\alpha$ & $\lambda_{s}$ & $\kappa_{r}$ & $r_{\epsilon}$ & R\\
\hline
$-6^{0}$ & $6^{0}$ & $-6^{0}$ & $91^{0}$ & $1,2 mm$ & $0,02 mm$\\
\hline
\end{tabular}
\caption{Geometrical characteristics of the tool}
	\label{tabl-1}
\end{table}

\section{Cutting torsor actions}
\label{sec:3}

\subsection{Tests}
\label{sec:Tests}

The experiments are performed within a framework similar to the one described in Cahuc et al., \cite{cahuc-AA-battaglia-01}. For each test, the complete torsor of the mechanical actions is measured using the six-component dynamometer according to the method initiated in Toulouse \cite{toulouse-98}, developed by Cou\'etard \cite{couetard-00a} and used in several occasions \cite{couetard-A-darnis-01}, \cite{cahuc-AA-battaglia-01}, \cite{darnis-A-couetard-00}, \cite{laheurte-AA-battaglia-03}, \cite{laheurte-A-cahuc-02}. These mechanical actions are evaluated for two depths of cut: ap = 2 and 5 mm; and for four feed rates: f = 0.05; 0.0625; 0.075  and 0.1 mm/rev. The six-component dynamometer gives the instantaneous values of all the torque cutting components in the three-dimensional space (\vec{x}, \vec{y}, \vec{z}) related to the machine tool (Fig.~\ref{fig1}). Measurements are performed in O', which is the center of the six-component dynamometer transducer. Then, they are transported to the tool point O via the moment transport classical relations. Measurement  uncertainties of the six-component dynamometer are: $±4\%$ for the forces components and $±6\%$ for the moment components.

To ensure the results repetitivity, for each configuration specified above, 4 series of measurements were carried out. For each series of measures, we analyzed the results over 15 seconds of recording. With 44 points of measurements recorded per spin round of the piece, the median values discussed thereafter correspond finally to the average of 7,590 measurements.

\subsection{Resultant of the cutting actions analysis}
\label{sec:ResultantOfTheCuttingActionsAnalyzes}

\begin{figure}[htbp]
	\centering
		\includegraphics[width=0.40\textwidth]{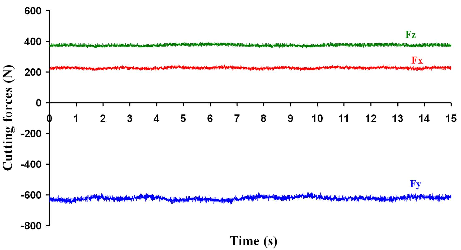}
	\caption{Signals related to the resultant components of cutting forces following the three (\vec{x}, \vec{y}, \vec{z}) cutting directions; test case using parameters ap = 2 mm, f = 0.1 mm/rev, and N = 690 rpm}
	\label{fig4}
\end{figure}

\begin{figure}[htbp]
	\centering
		\includegraphics[width=0.40\textwidth]{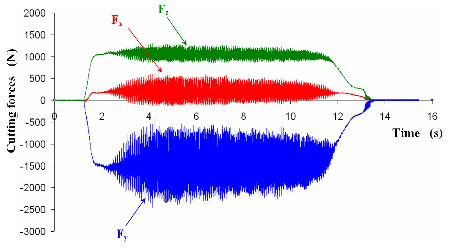}
	\caption{Signals related to the resultant components of cutting forces following the three (\vec{x}, \vec{y}, \vec{z}) cutting directions; test case using parameters ap = 5mm, f = 0.1 mm/rev, and N = 690 rpm}
	\label{fig5}
\end{figure}

For the four values feed rate f indicated above, two examples of forces resultant measurements applied to the tool tip point are presented: one of these for the stable case (quasi non-existent vibrations) ap = 2 mm (Fig.~\ref{fig4}), and another one for the unstable case (self-excited vibrations), ap = 5 mm, f = 0.1 mm/rev (Fig.~\ref{fig5}).

In the stable case (ap = 2 mm, f = 0.1 mm/rev), it appears that the force component amplitudes remain almost independent of the time parameter. Thus, the amplitude variation is limited to 1 or 2 N around their rated values, starting with 200 N for (F$_{x}$) 400 N for (F$_{z}$) and 600 N for (F$_{y}$). These variations are quite negligible and lower than uncertainties of measurements. Indeed, when the rated stress is reached, the component noticed as the lowest value is (F$_{x}$), while the highest in the absolute value is ($F_{y}$). Taking as reference the absolute value of ($F_{x}$), the following relation between these three components comes:

\begin{equation}
 \left|F_{x}\right| = \frac{\left|F_{z}\right|}{2} = \frac{\left|F_{y}\right|}{3}.
\end{equation}

In the unstable case, we observe that the force component on the cutting axis ($F_{y}$) has the most important average amplitude (1,500 N). It is also the most disturbed ($±1,000$ N), with oscillations between -2,450 N and -450 N. The force along the feed rate axis ($F_{z}$) has also an important average amplitude ($1,000$ N), but the oscillations are smaller: $±200$ N in absolute value and $±20\%$ in relative value. The efforts along the radial direction ($F_{x}$) are the weakest in terms of average amplitude (200 N), but also the most disturbed regarding the relative value ($±240$ N). These important oscillations are the tangible consequence of the contact tool/workpiece frequent ruptures and, thus, demonstrate the vibration and dynamical behaviour of the system workpiece-tool-machine \textbf{WTM}.

Finally, taking as reference the absolute value of ($F_{x}$), the following relation between the absolute values of these three components comes:

\begin{equation}
 \left|F_{x}\right| = \frac{\left|F_{z}\right|}{5} = \frac{\left|F_{y}\right|}{7.5} .
\end{equation}

In other words, whatever the depth of cut ap, we have the following order relation between the absolute values of the cutting actions components:

\begin{equation}
        |Fx| \leq |Fz| \leq |Fy|.						   
\end{equation}

More precisely, in the case of turning studied here, for the depth of cut ap = 2 mm (stable case) or ap = 5 mm (unstable case) we have, with the errors of experiments close the following relation between the absolute values of the cutting actions components:

\begin{equation}
     \left|F_{x}\right| = \frac{\left|F_{z}\right|}{ap} = \frac{\left|F_{y}\right|}{1.5ap} .				 
\end{equation}

\begin{figure}[htbp]
	\centering
			\includegraphics[width=0.48\textwidth]{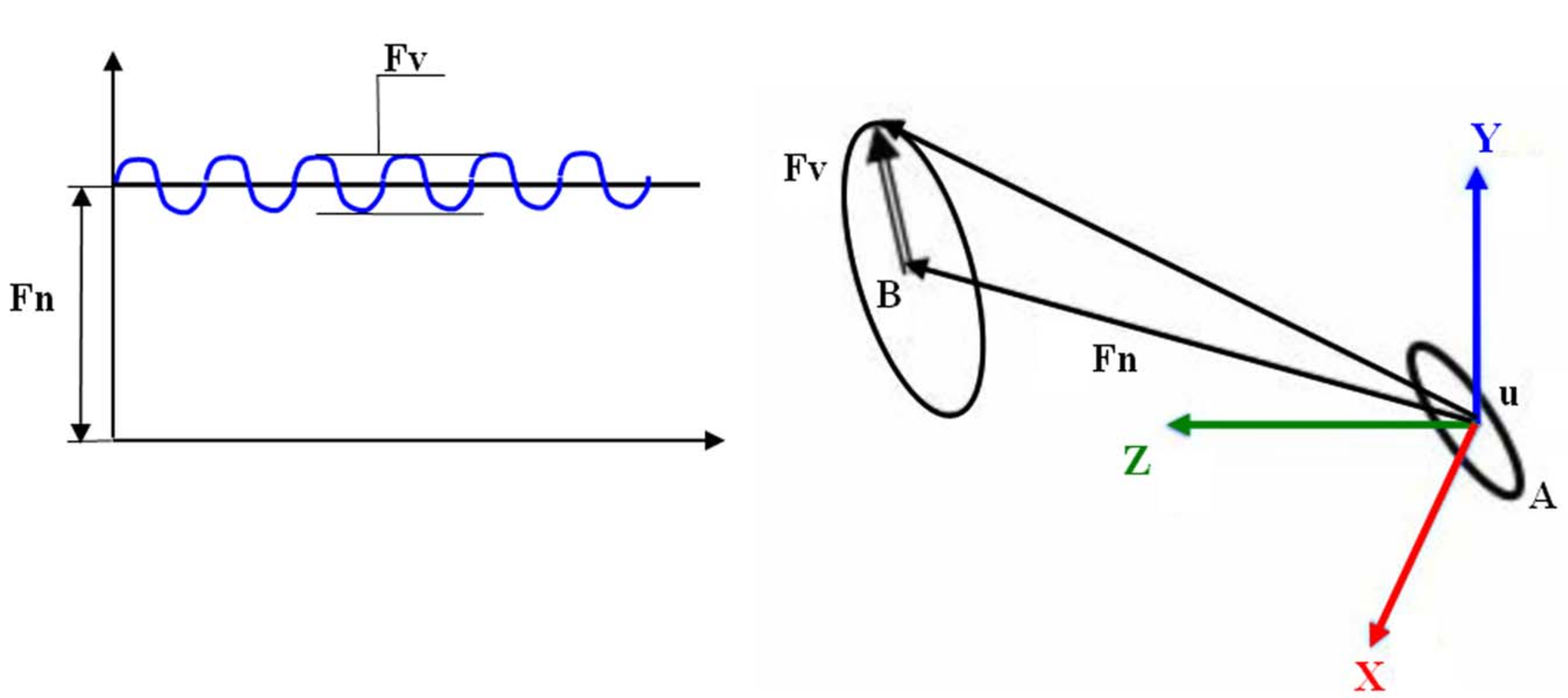}
	\caption{Cutting force F$_{v}$ evolution around the nominal value F$_{n}$.}
	\label{fig6}
\end{figure}

We find that the cutting actions are proportional to the depth of cut. This result is in agreement with \cite{benardos-A-vosniakos-06}, \cite{dimla-04}. In the unstable case (ap = 5mm, f = 0.1 mm/rev) the force resultant component detailed highlights a plan in which a variable cutting forces F$_{v}$ moves around a rated value F$_{n}$ \cite{bisu-AAA-cahuc-09}. This variable force is an oscillating action (Fig.~\ref{fig6}) that generates tool tip displacements and maintains the vibrations of elastic system block-tool \textbf{BT}. Thus, the cutting force variable and the self-excited vibrations of elastic \textbf{WTM} system are interactive, in agreement with research work \cite{lian-A-huang-07}, \cite{koenigsberger-tlusty-70}, \cite{dimla-04}, \cite{stawell-00}.

\subsection{Study of the moments to the tool point}
\label{sec:StudyOfTheMomentsToTheToolPoint}

\begin{figure}[htbp]
	\centering
		\includegraphics[width=0.48\textwidth]{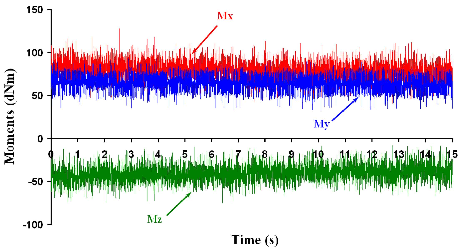}
	\caption{The moment components time signals that act on the tool tip following the three directions of space machine in the stable case ap = 2 mm, f = 0.1 mm/rev and N = 690 rpm}
	\label{fig7}
\end{figure}

\begin{figure}[htbp]
	\centering
		\includegraphics[width=0.48\textwidth]{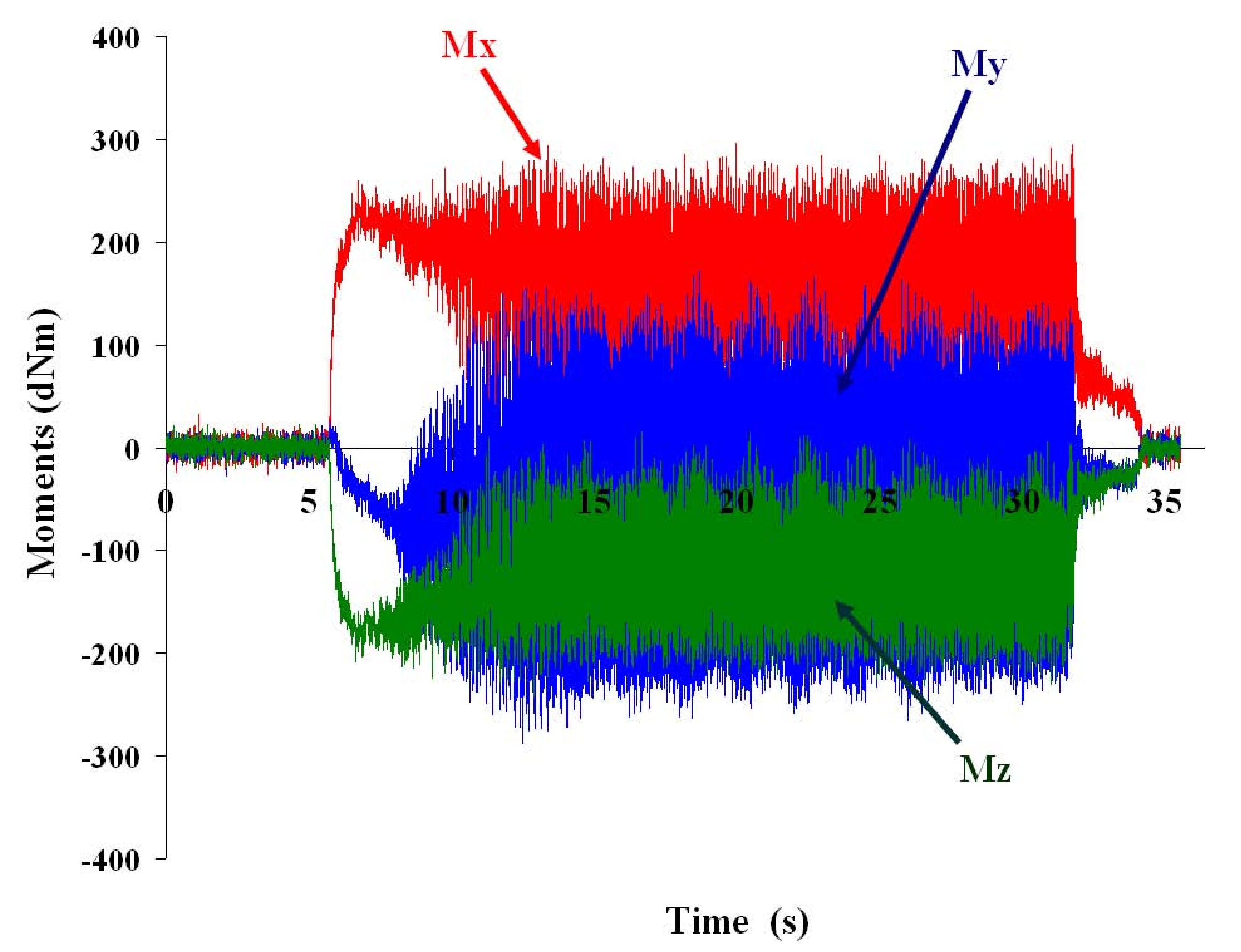}
	\caption{The moment components time signals that act on the tool tip following the three directions of space machine in the unstable case ap = 5 mm, f = 0.1 mm/rev and N = 690 rpm}
	\label{fig8}
\end{figure}

Like for the torsor resultant, we give here two examples of statements of the moment components at the tool point in the machine frame (Fig.~\ref{fig1}). The stable case is first presented, see Fig.~\ref{fig7} (without vibrations ap = 2 mm). Then the unstable case is shown in Fig.~\ref{fig8} (with vibrations ap = 5 mm).

Like for the resultants in the stable case (ap = 2 mm, f = 0.1 mm/rev), the moment components average values are slightly disturbed at the tool tip point (quasi-constants except for some dN.m) though in the unstable case (ap = 5 mm, f = 0.1 mm/rev), the moment components average values are very disturbed at the tool tip point. However, we can note that the moment components at the tool tip point are more disturbed than their equivalents regarding the resultant. Thus, the moment components seem more sensitive to the self-excited vibrations than the force components applied. Thus, the follow-up of those could be a good mean of precociously detecting the existence of regenerative vibrations.

The analysis of the results shown on Fig.~\ref{fig7} allows to establish that, among the three moment components, the average value of the component along the axis feed rate is the lowest (M$_{oz}$ = - 40 dN.m). It is the most disturbed in relative value ($±20$ dN.m i.e. 50\%). The absolute value of the component along the axis of cut plays the role of pivot (M$_{oy}$ = 60 dN.m) with a weaker disturbance in absolute value ($±16$ dN.m i.e. 30\%). The highest component is along the radial axis (M$_{ox}$ = 80 dN.m) with a weaker disturbance in absolute value ($±20$ dN.m i.e. 25\%). Taking as reference the average values of the component M$_{oz}$, it comes : 

\begin{equation}
     \left|M_{oz}\right| = \frac{\left|M_{oy}\right|}{1.5} = \frac{\left|M_{ox}\right|} {2}, (ap = 2 mm). 					 
\end{equation}

The analysis of the results show on Fig.~\ref{fig8} allows to establish that among the three moment components, the average value of the component along the cutting axis is the lowest (M$_{oy}$ = - 100 dN.m). It is also the most disturbed ($±280$ dN.m i.e. variations of about 280\%!). The component of the torque along the radial axis (\vec{x}) is continuously positive M$_{ox}$ = 170 dN.m ($±150$ dN.m); it is thus the most raised moment component and, proportionally, the least disturbed ($±90$ \%). The M$_{oz}$ moment component along the feed rate direction is continuously negative, with an average value about - 135 dN.m in absolute value, the highest, but less disturbed than M$_{oy}$ (with oscillations of $±140$ dN.m i.e. only $±105$ \%). Finally, for this depth of cut, we have the following relation between the absolute values of the moment components:

\begin{equation}
     \left|M_{oy}\right| = \frac{\left|M_{oz}\right|}{1.4} = \frac{\left|M_{ox}\right|}{1.7}, (ap= 5 mm), 					 
\end{equation}

with a module of average torque at the tool tip point about $\left|M_{o}\right|$  = 275 dN.m. The comparison of Eq. (2) and Eq. (6), shows that the role of \vec{x} and \vec{y} axes is reversed for ap = 5 mm. This allows to think that the transport of the moments at the tool tip point has a major effect. Moreover, whatever the value of ap (= 2 mm; 5 mm), M$_{ox}$ modulus keeps the highest value in the turning cases here considered. It is also the only positive and slightly decreasing component when the depth of cut increases. Furthermore, one notes that the components M$_{oy}$ and M$_{oz}$ are continuously negative when the depth of cut increases. One notes also that the component M$_{oy}$ is decreasing while the component M$_{oz}$ is increasing. The modulus of the moments component at the tool tip point, increases with the depth of cut, like the one of the resultant. 

In the same way, for ap = 5 mm, the general order relation (3) for the resultant is replaced for the moment at the tool tip point by:

\begin{equation}
     \left|M_{oy}\right| \leq \left|M_{oz}\right| \leq \left|M_{ox}\right|.					 
\end{equation}

The \vec{x} and \vec{y} axes positions are also reversed regarding the resultant and moment at the tool tip point components. This can be allotted to the moments transport at the tool tip point.
 
\begin{figure}[htbp]
	\centering
		\includegraphics[width=0.49\textwidth]{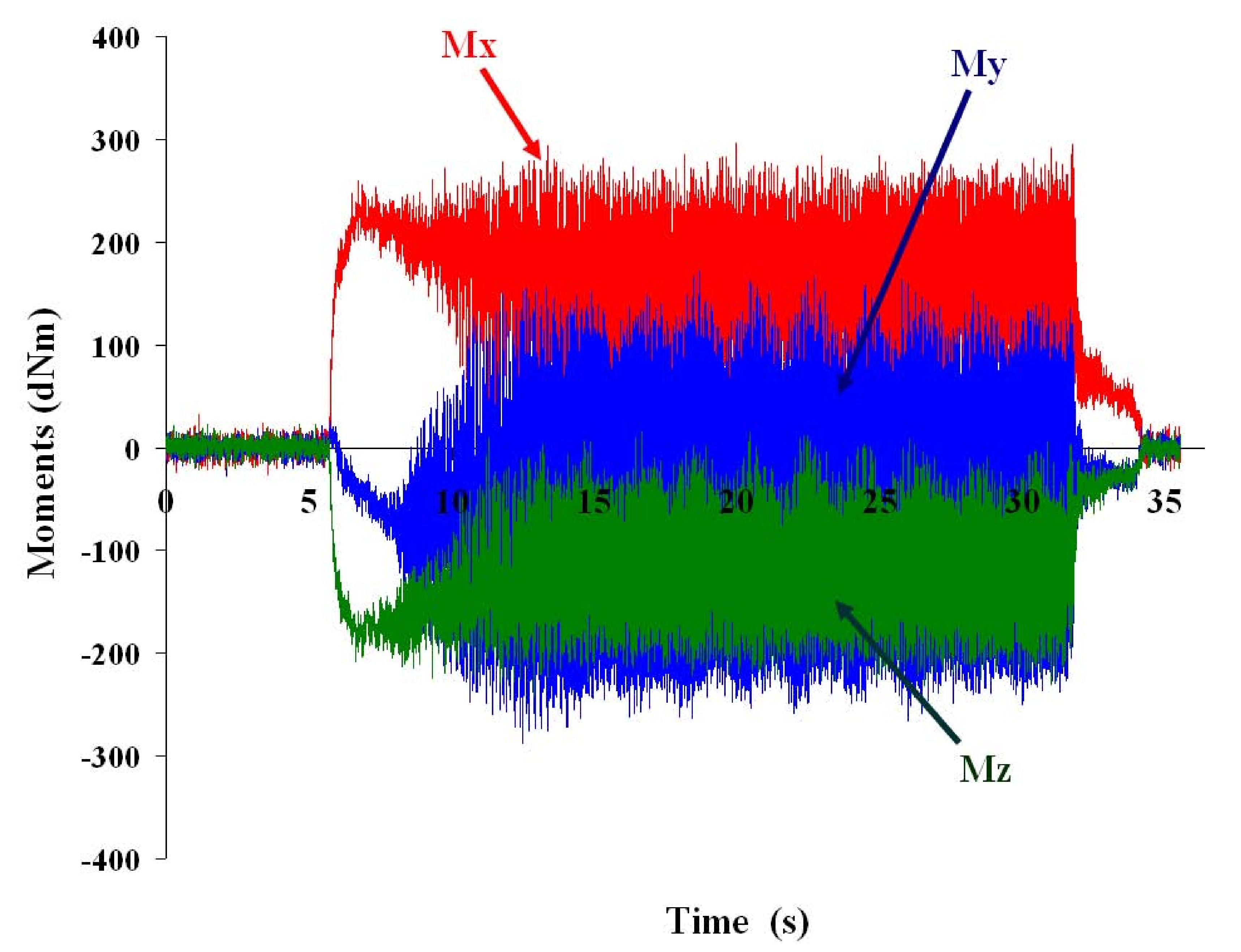}
	\caption{Evolution of the moment components at the tool tip point in the tool tip reference system for ap = 5 mm}
	\label{fig9}
\end{figure}

It is always important to know the influence of the vibrations on the behavior of the moments at the tool tip point. So, we remain in the case of the strong vibrations (ap = 5 mm) to examine their influence when the feed rate increases. An example of the behavior of the moment components average values at the tool tip point is described in figure \ref{fig9}. The following conclusions are obtained (ap = 5 mm, f variable):

\begin{itemize}

\item[-] the validity of the order relation of (7) is extended to the case of a variable feed rate,

\item[-] the component M$_{oz} $ which consumes the mechanical power goes through a light extremum around feed rate f = 0.0625 mm/rev,

\item[-] the component M$_{oy} $ varies like the component M$_{oz} $, but in more accentuated way, and also goes through a extremum more marked around f = 0.0625 mm/rev,

\item[-] the component M$_{ox} $ is quasi-constant and continuously negative,

\item[-] finally the results indicated above are more accentuated in the tool reference system than in the machine tool reference system.

\end{itemize}

These results show that it is better to follow the moments evolution in the tool reference system than this one of the machine tool. Let us see now in the following section the influence of the vibrations on the central axis of the mechanical actions torsor.

\section{Moments analysis at central axis} 
\label{sec:MomentsAnalysisToCentralAxis}

\subsection{Central axis} 
\label{sec:CentralAxis}

It is well-known that, it is possible to associate a central axis, to any torsor (except the torsor of pure moment),  which is the single object calculated starting from the six torsor components \cite{brousse-73}. A torsor $\left[A\right]_{O}$ in a point O is composed of the resultant forces \vec{R} and the resulting moment \vec{M}$_{O}$ :

\begin{eqnarray}
	[A]_{O} = \left\{
	\begin{array}{c}
	\vec{R} ,\\
  \vec{M_{O} .}
\end{array}\right.
\end{eqnarray}

\begin{figure}[htbp]
	\centering
		\includegraphics[width=0.48\textwidth]{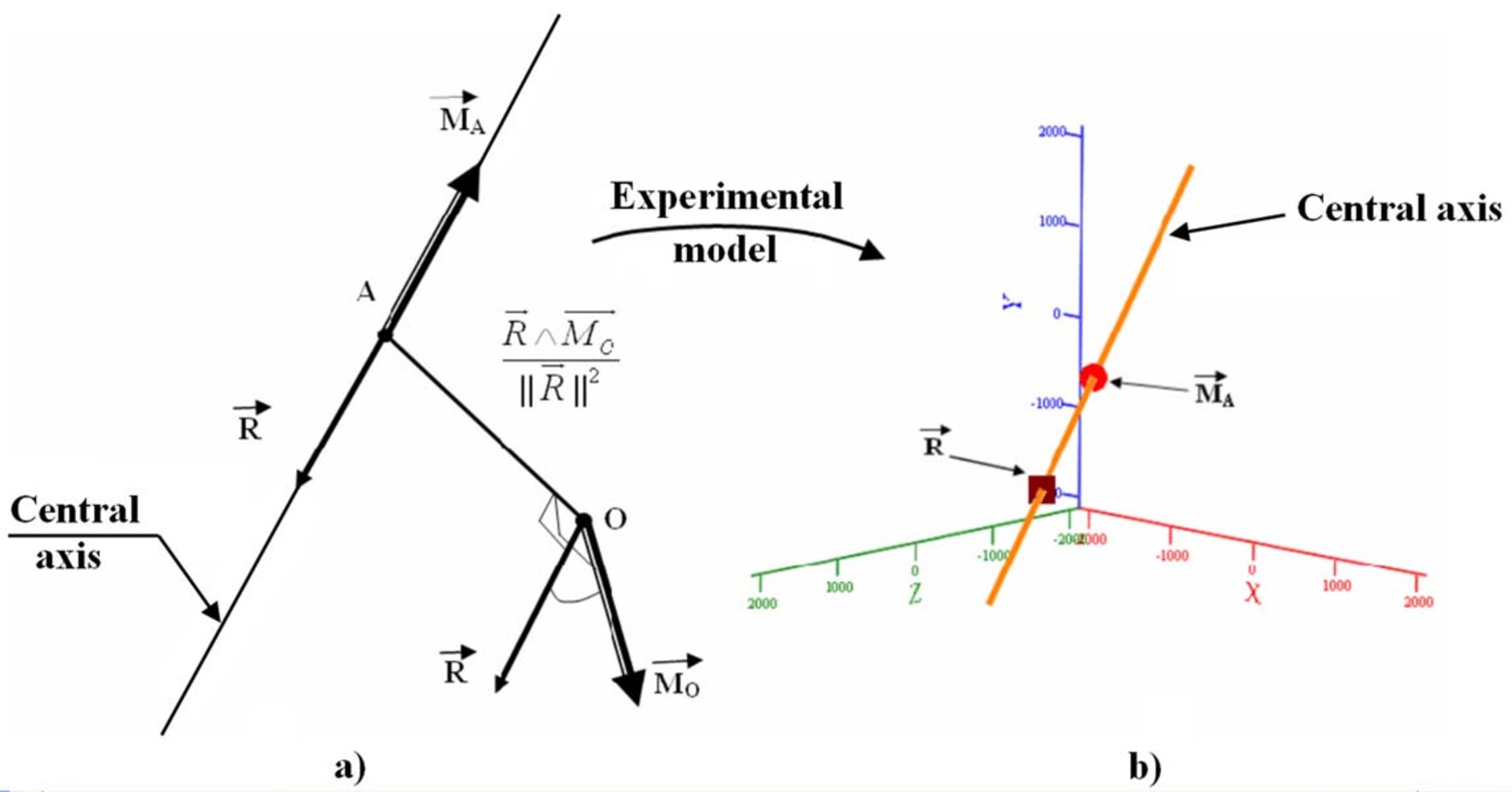}
	\caption{ Central axis representation (a) and of the colinearity between vector sum \vec{R} and minimum moment \vec{M}$_{A}$ on central axis (b)}
	\label{fig10}
\end{figure}

The central axis is the straight line classically defined by:

\begin{equation}
	\vec{OA}=\frac{\vec{R}\wedge\vec{M}_{O}}{\left|\vec{R}^{2}\right|}+\lambda\vec{R} ,
\end{equation}

where O is the point where the mechanical actions torsor was moved (here, the tool tip) and A the current point describing the central axis. Thus, \vec{OA} is the vector associated with the bi-point [O, A] (Fig.~\ref{fig10}).

This line (Fig.~\ref{fig10}a) corresponds to the geometric points where the mechanical actions moment torsor is minimal. The central axis calculation consists in determining the points assembly (a line) where the torsor can be expressed along a slide block (straight line direction) and the pure moment (or torque) \cite{brousse-73}.

\begin{figure}[htbp]
	\centering
		\includegraphics[width=0.48\textwidth]{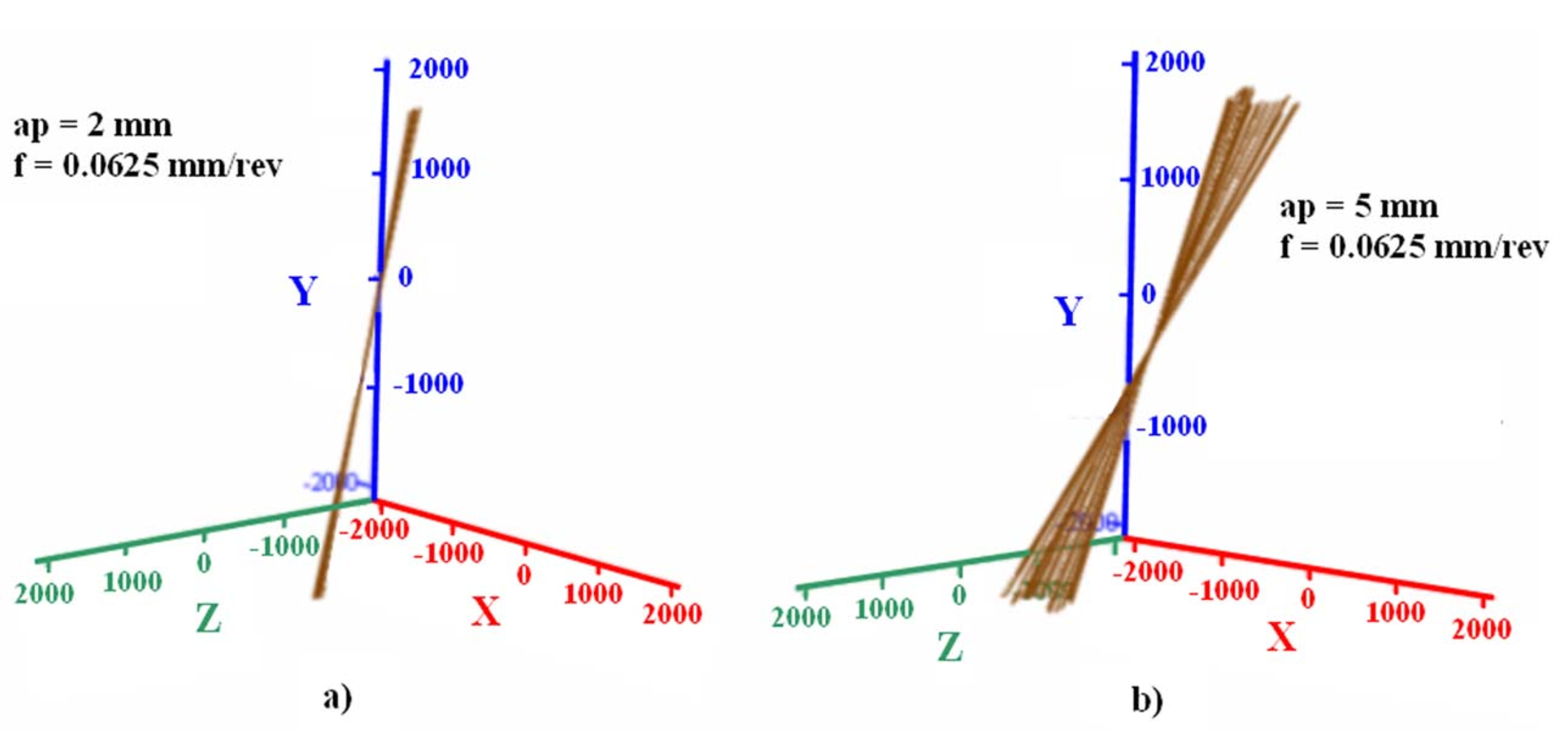}
	\caption{Central axes representation obtained for 68 spin rounds of the workpiece feed rate f = 0.0625 mm/rev; a) stable process ap = 2 mm; b) unstable process ap = 5 mm}
	\label{fig11}
\end{figure}

The central axis is also the point where the resultant cutting force is colinear with the minimum mechanical moment (pure torque). The test results enable us to check, for each measurement point, if there is colinearity between the resultant cutting force \vec{R} and moment \vec{M}$_{A}$ calculated at a point of the central axis (Fig.~\ref{fig10}b). The meticulous examination of the six mechanical action torsor components shows that the forces and the moment average values are not null. For each point of measurements, the central axis is calculated in the stable (Fig.~\ref{fig11}a) and unstable modes (Fig.~\ref{fig11}b). In any rigour, the case ap = 2 mm and f = 0.1 mm/rev should be described as quasi-stable movement because the vibrations exist, but their amplitudes are very small -of the order of micrometers- thus, quasi null compared to the other studied cases. Considering the cutting depth value ap = 5 mm and f = 0.0625 mm/rev, the recorded amplitude was 10 times more important.

The central axes beam is larger in unstable case, and is also more tilted compared with the normal axis on the plane (x, y). The self-excited vibrations, due to the variable moment generation, can explain this central axis dispersion. 

\subsection{Moments analysis}
\label{sec:5}

The minimum moment (pure torque) $\vec{M}_{A}$ is obtained by transporting the moment from tool tip at the central axis. Constant and variable parts are deduced from the moment values at the central axis. Like for the efforts, the variable part is due to the self-excited vibrations, as revealed below.

The moments contribution on the areas of contact tool-workpiece-chip is expressed using this decomposition. The observations resulting from the analysis show that the tool vibrations generate rotations; cause variations of contact; and, thus, generate variable moments, confirming the efforts analysis detailed in Section~\ref{sec:3}. {Let us express the moments along the three axes of the machine tool: swivel moment in the \vec{y} direction and the two rolling moments along \vec{x} and \vec{z} directions.

For the unstable case, characterized by ap = 5 mm and f = 0.1 mm/rev, the analysis of the cutting torque components average values at the central axis (subscripted by the letter a) allows to establish that among the three moment components, the average value of the component along the radial axis is (in absolute value) the lowest (M$_{ax}$ = - 4 dN.m). It is very disturbed in relative value ($±24$ dN.m  i.e. 600\%). The absolute value of the component along the cutting axis is the highest (M$_{ay}$ = - 34 dN.m) with a weaker disturbance in absolute value ($±160$ dN.m i. e. 470\%). The component along the feed rate axis is medium (M$_{az}$ = 8 dN.m). It is the most disturbed in relative value ($±110$ dN.m i. e. 1,300\%). Thus, one has, by the errors of measurement, the following relation between the absolute values of the three torque components:
 
\begin{equation}
     \left|M_{ax}\right| = \frac{\left|M_{az}\right|}{2} = \frac{\left|M_{ay}\right|}{8}, (ap = 5 mm).					 
\end{equation}

The confrontation of Eq (6) and (10) shows the interest of transporting the moment from the tool tip point  to central axis. It is also interesting to note that the role of pivot of \vec{z} axis is preserved. The role of \vec{x} and \vec{y} axes is reversed here. Thus, the order relation (7) is replaced by:

\begin{equation}
     \left|M_{ax}\right| \leq \left|M_{az}\right| \leq \left|M_{ay}\right|.					 
\end{equation}

At the central axis, the modulus of the torque components and those of the forces are always similar. Thus, the torque along the feed rate axis \vec{z} is framed by the component on the radial axis \vec{x} which is the weakest. The strongest is on the cutting axis \vec{y}. It is as also interesting to note as the modulus of the average torque of cut at the central axis is worth to $\left|M_{a}\right|$ = 90 dN.m, either 3 times less than at the tool point.

This result encourages us to study again the influence of the vibrations on the moments evolution at the central axis when the feed rate varies. Let us consider the case of the strong vibrations (ap = 5 mm) to examine their influence when the feed rate increases.  An example of the behavior of the average values of the moment components at the central axis in the tool reference system is given at figure~\ref{fig12}. We note that the evolution of the moments components at the central axis is very different from those recorded at the tool tip figure~\ref{fig9}.

\begin{figure}
	\centering
		\includegraphics[width=0.49\textwidth]{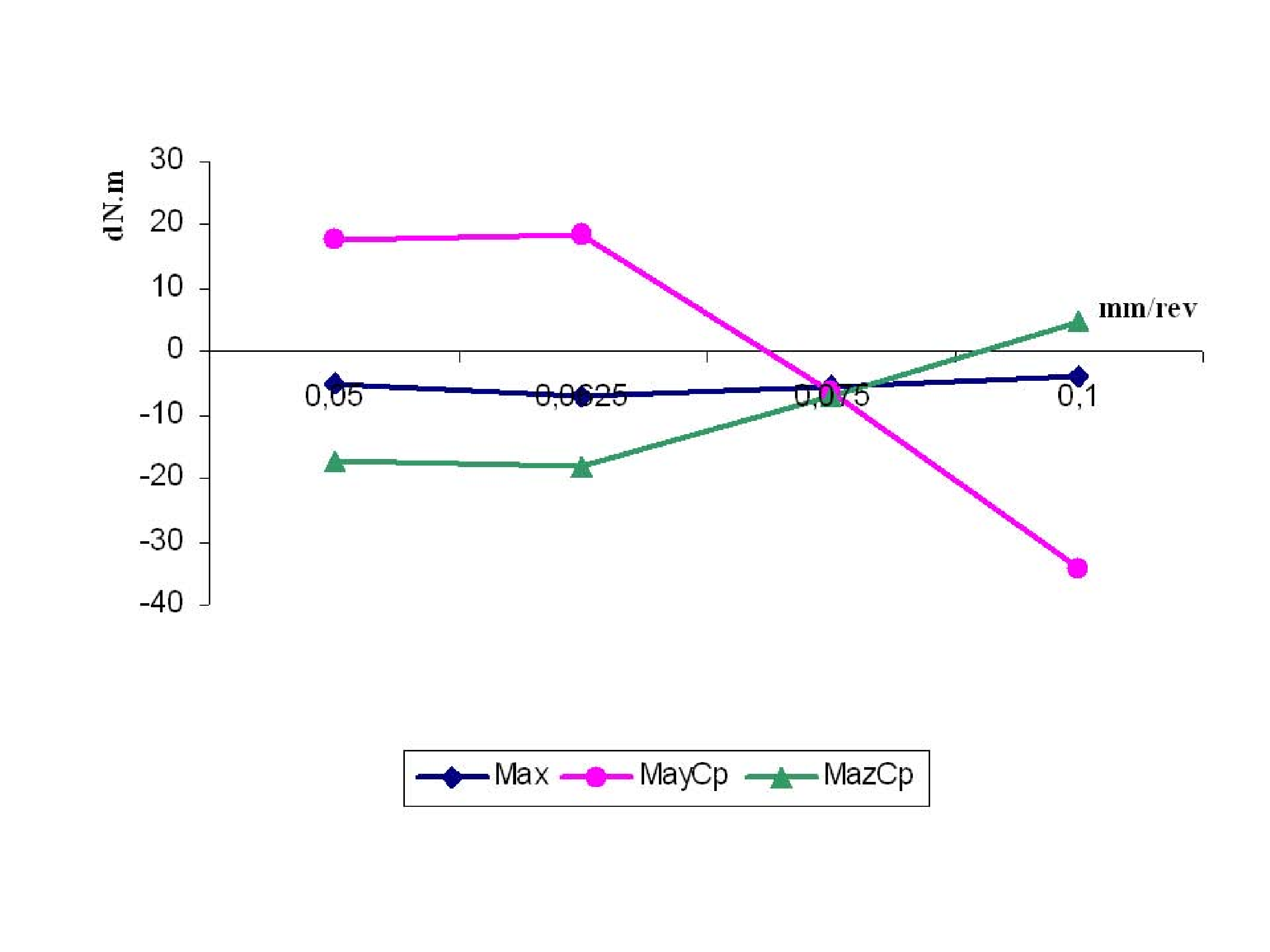}
	\caption{Evolution of the moments components at the central axis in the tool tip reference system for ap = 5mm.}
	\label{fig12}
\end{figure}

The examination of the moment components at the central axis is enriching. We obtain the following conclusions (ap = 5 mm, f variable):

\begin{itemize}

\item[-] In average, all the components of the moments at the central axis have the same value for f= 0.075 mm/rev.

\item[-] Only the M$_{ax}$ component is quasi-constant and keep the same sign (negative). It is the smallest component. These properties are opposite to the ones noticed at the tool tip point. It is also a tangible consequence of the forces influence in the transport of the moments. This influence comes to disturb the analysis of the moments and to hide the consequences which one can deduce.

\item[-] The M$_{az}$ component which consumes the mechanical energy is initially quasi-constant and negative till f = 0.0625 mm/rev, then strongly increasing after. It becomes positive after the value f = 0.075 mm/rev.

\item[-] Conversely, the M$_{ay}$ component is initially positive and quasi-constant till f = 0.0625 mm/rev, then strongly decreasing after and becomes negative after the value f = 0.075 mm/rev.

\item[-] Finally the results indicated above are accentuated in the tool reference system than in the machine tool one,

\item[-] For variable feed rate (ap = 5 mm), one notes that the torque modulus is increasing with feed rate as for the resultant cutting force, on the central axis.

\item[-] The single order relation (11) is replaced by the two following:

\begin{equation}
     \left|M_{ax}\right| \leq \left|M_{ay}\right| \leq \left|M_{az}\right|	\text{ (if, $ f \leq 0.0625 mm/rev)$},				 
\end{equation}

\begin{equation}
     \left|M_{ax}\right| \leq \left|M_{az}\right| \leq \left|M_{ay}\right|	\text{(if, $ f > 0.0625 mm/rev)$}.				 
\end{equation}

\item[-] These two relations are associated to a modification of the localization of the rolling moment in the (y, z) tool plan (and in agreement with figure~\ref{fig12}). This one passes from the fourth quadrant of the plan (y, z) for f $\leq$ 0.0625 mm/rev to the second quadrant for f $>$ 0.0625 mm/rev. This change of the bearing moment orientation is associated to a modification in the ejection of the chip. This one is evacuated towards the left of the \vec{x} axis (f $\leq$ 0.0625 mm/rev) then towards the right of the x axis (f $>$ 0.0625 mm/rev). This analysis of the results presented at figure~\ref{fig12} is confirmed by the examination of the corresponding recording videos.

\end{itemize}

The results show that it is better to study the moments evolution in the tool reference system than in the machine tool system. It is also obvious that the analysis of the mechanical actions moment must be performed at the central axis.

\section{Conclusion}
\label{conclusion}

Experimental procedures here developed allowed to determine the elements required for a rigorous analysis of the influence of tool geometry, its displacement and the evolution of contacts tool/workpiece and tool/chip on the machined surface. These experimental results allowed to establish a vectorial decomposition of actions analysing the resultant of applied actions torsor during a turning process. Thus, it is demonstrated that the cutting force evolves around a constant rated value. This variable force moves in a plan tilted compared to the machine spindle. 

Furthermore, a really innovative study of the moments at the tool tip point, and at the central axis of the mechanical action torsor, is presented.  For the resultant like for the moment, at the tool tip point or the central axis, the modulus of these elements is increasing with the feed rate. If one compares the average values of the components modulus action with those of the moments at the tool tip point an order relation clearly appears. In particular one notes that the roles of \vec{x} and \vec{y} axes are inversed, and \vec{z} axis plays the role of "pivot". On the other hand, at the central axis there does not exist a unic single order relation but two. These two relations corresponds to a modification of the localization of the rolling moment in the (y, z) tool plan. This one goes from the fourth quadrant of 
the plan (y, z) for f $\leq$ 0.0625 mm/rev to the second quadrant for f $>$ 0.0625 mm/rev. This change of the rolling moment orientation is associated to a modification in the ejection of the chip. This one is evacuated towards the left of the \vec{x} axis (f $\leq$ 0.0625 mm/rev) then towards the right of the \vec{x} axis (f $>$ 0.0625 mm/rev). This analysis of the results presented in figure~\ref{fig12} and is confirmed by the examination of the corresponding recording videos.

Furthermore, this review allows to better understand the results obtained in (\cite{bisu-AAA-cahuc-09}, Fig. 13). Indeed, it was noticed that the place of moments has a shape of buttonhole mainly inscribed in a plan. Its tips come slightly out of this plan along a small segment. This particular buttoncoole shape reveals that the moment is passing from the second to the fourth quadrant (see also \cite{bisu-AAA-cahuc-09}, Fig. 16).

Only the moment modulus along the radial axis remains the weakest when the feed rate increases. This property is common with the component modulus of the same row of the resultant. For the modulus of the other components of the moments at the central axis, their relative position depends on the feed rate value. Moreover, one notes that the moment components at the central axis are subjected to oscillations much more important than these same components at the tool tip point. This shows that the moments components at the central axis are particularly sensitive to the disturbances of machining, here the self-excited vibrations. Thus, the follow-up of these moments evolution at the central axis allow to follow, a priori, the surface quality of the machined piece. The more important the amplitudes of vibrations of tools are, the poorer the surface quality of the piece manufactured is. This suggests one should look further into the study of the cutting moments, especially, at the central axis. More precisely, all these remarks show that it is particularly important to be placed under cutting conditions such that the central axis of the torsor passes by the cutting line (or so not with more close possible of the cutting line). Indeed, such a configuration corresponds to the cutting moments minimum. This situation leads then to the minimum friction conditions supporting of the minimal rises in temperature, and thus also a minimum tool wear, since all these elements are dependant and increasing with the applied forces \cite{toh-04}, \cite{lalwani-A-jain-08}.

\begin{acknowledgements}
The authors would like to thank the CNRS (Centre National de la Recherche Scientifique UMR 5469) for the financial support of this project. 

\end{acknowledgements}

\end{document}